\documentclass[aps,prl,preprint,floatfix,superscriptaddress,showpacs]{revtex4-1}
\usepackage{graphicx}
\usepackage{amssymb,amsmath}

\usepackage{nicefrac}

\begin{document}
\widetext

\title{Nuclear resonant surface diffraction}

\author{Kai Schlage} \affiliation{Deutsches Elektronen-Synchrotron DESY, Notkestr. 85, 22607 Hamburg, Germany}
\author{Liudmila Dzemiantsova}
\affiliation{Deutsches Elektronen-Synchrotron DESY, Notkestr. 85, 22607 Hamburg, Germany}
\affiliation {The Hamburg Centre for Ultrafast Imaging, Luruper Chausee 149, 22671 Hamburg}
\author{Lars Bocklage} 
\affiliation{Deutsches Elektronen-Synchrotron DESY, Notkestr. 85, 22607 Hamburg, Germany}
\affiliation {The Hamburg Centre for Ultrafast Imaging, Luruper Chausee 149, 22671 Hamburg}
\author{Hans-Christian Wille} 
\affiliation{Deutsches Elektronen-Synchrotron DESY, Notkestr. 85, 22607 Hamburg, Germany}
\author{Matthias Pues}
\affiliation{Institut f\"ur Nanostruktur- und Festk\"orperphysik, Jungiusstr. 11, 20355 Hamburg}
\author{Guido Meier} 
\affiliation{Deutsches Elektronen-Synchrotron DESY, Notkestr. 85, 22607 Hamburg, Germany}
\affiliation {The Hamburg Centre for Ultrafast Imaging, Luruper Chaussee 149, 22671 Hamburg}
\affiliation {Max Planck Institute for the Structure and Dynamics of Matter, Luruper Chaussee 149, 22761 Hamburg}
\author{Ralf R\"ohlsberger} 
\affiliation{Deutsches Elektronen-Synchrotron DESY, Notkestr. 85, 22607 Hamburg, Germany}
\affiliation {The Hamburg Centre for Ultrafast Imaging, Luruper Chaussee 149, 22671 Hamburg}
\vskip 0.25cm
\date{\today}

\begin{abstract}
Nuclear resonant x-ray diffraction in grazing incidence geometry is used to determine the lateral magnetic configuration in a one-dimensional lattice of ferromagnetic nanostripes. During magnetic reversal, strong nuclear superstructure diffraction peaks appear in addition to the electronic ones due to an antiferromagnetic order in the nanostripe lattice. We show that the analysis of the angular distribution of the resonantly diffracted x-rays together with the time-dependence of the coherently diffracted nuclear signal reveals surface spin structures with very high sensitivity. This novel scattering technique provides a unique access to laterally correlated spin configurations in magnetically ordered nanostructures and, in perspective, also to their dynamics.
\end{abstract}

\maketitle
\begin{figure}[t] 
	\centering
\hspace*{-0.3cm}
\includegraphics[scale=0.16]{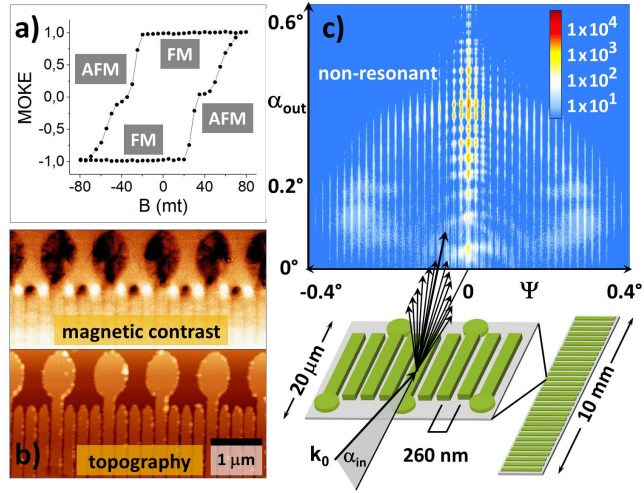}
\caption{\label{setup} 
Sample description and scattering geometry. (a) Hysteresis curve recorded via the magneto-optical Kerr effect (MOKE) shows that a permalloy nanostripe pattern in pad geometry can be switched between a ferromagnetic and an antiferromagnetic state. (b) An atomic force micrograph visualizes the pad structure which alternatingly reduces the coercivity of every second nanostripe. A magnetic force micrograph confirms a local antiferromagnetic configuration during magnetic reversal on the nanoscale. (c) The non-resonant 2D x-ray diffraction pattern is characterized by a large number of diffraction maxima indicating the high structual quality of the nanostripe lattice.}
\end{figure}
\begin{figure}[t]
\hspace*{-0.4cm}
 \vspace*{-0.3cm}
\includegraphics[scale=0.20]{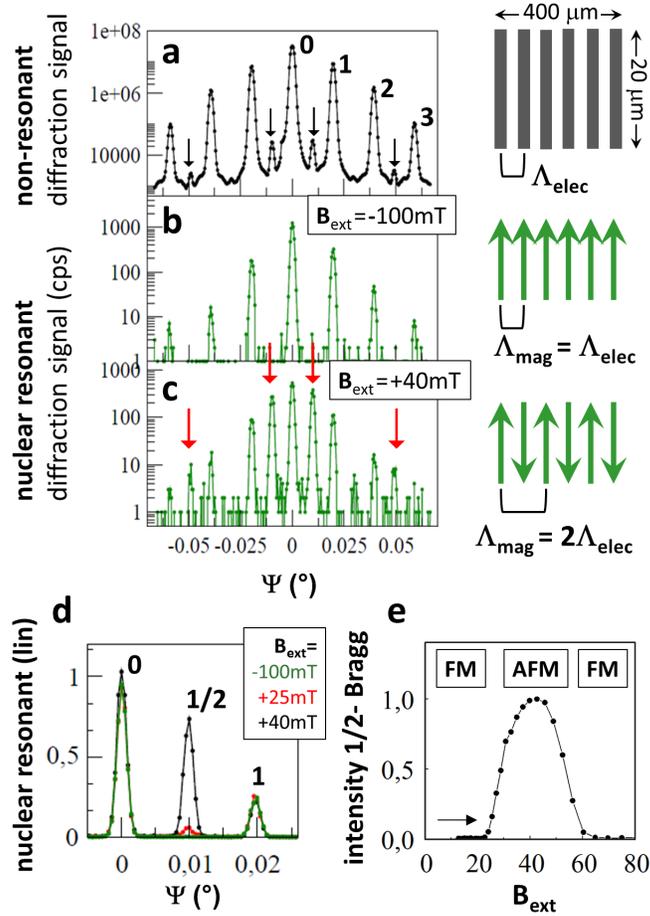}
\caption{\label{fig:epsart} Non-resonant and nuclear resonant out-of-plane diffraction patterns at the position of the specular reflected beam ($\alpha_{\text{out}}=\alpha_{\text{in}}=0.2^\circ$). (a) Electronic diffraction pattern. The seperation of the electronic diffraction maxima results from the structural periodicity of the nanostripe pattern with a lateral correlation length $\Lambda_{\text{elec}}$ of 260 nm. The weak diffraction peaks (black arrows) result from the additional pad structure with a doubled lateral spacing. (b) Nuclear resonant diffraction pattern of the magnetically saturated sample. The angular distribution of the nuclear diffraction pattern reproduces the electronic ones with a relative intensity of around 10$^{-4}$. (c) During magnetic reversal additional strong superstructure diffraction peaks appear (red arrows). A lateral magnetic correlation length $\Lambda_{\text{mag}}$ can be identified which is doubled compared to the electronic one. This confirms an antiferromagnetic configuration in the lattice. (d) Evolution of the nuclear resonant diffrcation pattern during magnetization reversal. (e) Intensity of the 1/2-order Bragg peak as a function of the external field strenght.}
\end{figure}
\begin{figure*}
\vspace*{-0.1cm}
\hspace*{1.0cm}
\includegraphics[scale=0.19]{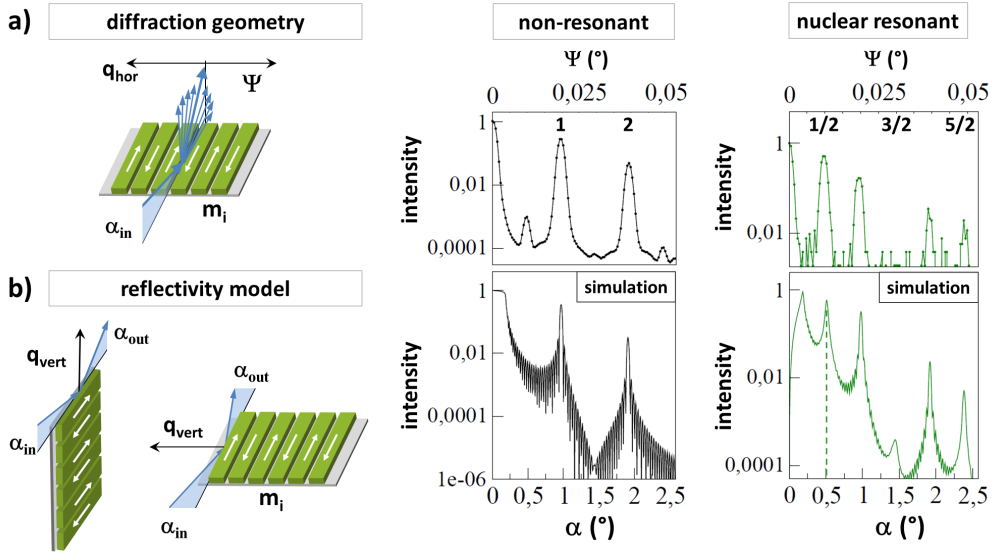}
\caption{\label{fig:wide} Analogy between nuclear resonant diffraction from a surface grating and nuclear reflectometry at a periodic multilayer. (a) Diffraction geometry and experimental diffraction pattern from the nanostripe lattice in the antiferromagnetic configuration. (b) Sketch of the reflectivity model used to simulate the diffraction pattern. Calculated data sets reproduce the overall shape of the electronic and nuclear diffraction data and reveal the origin of the suppression of the second magnetic Bragg peak (3/2 Bragg). The suppression of this peak can be attributed to the ratio of the lateral magnetic correlation length to the width of the nanostripes which is close to 3:1. 
}
\end{figure*}
\begin{figure}[b]
\includegraphics[scale=0.17]{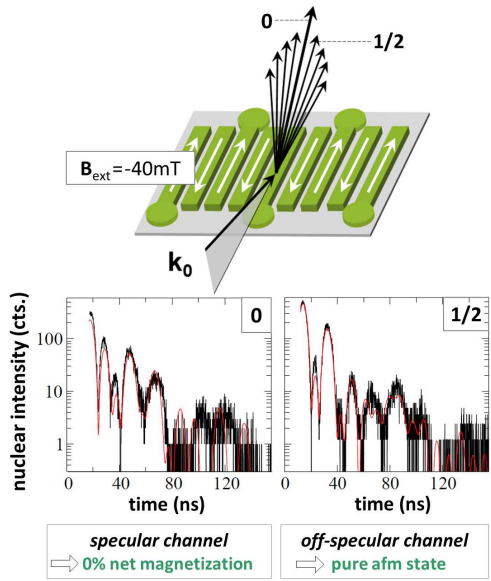}
\caption{\label{timespectra} Dependence of the coherent nuclear decay of the resonantly excited $^{57}$Fe nuclei on the selected scattering channel. The left data set shows a nuclear time spectrum detected in specular reflection together with a fit (red line) using the CONUSS program package for nuclear reflectometry \cite{Stur00}. A zero net magnetization of the nano stripe lattice can be identified. The time spectrum on the right is detected at the angular position of the first magnetic Bragg peak and can be evaluated using the reflectivity model presented in Fig. 3. An excellent agreement with the measured data is achieved if a pure antiferromagnetic configuration of the lattice is assumed. 
}
\end{figure}

Resonant magnetic x-ray scattering of synchrotron radiation \cite{Hann88, Gibbs, Lovesey1996} is an established tool to probe magnetism in crystalline systems, ultra-thin films and multilayer structures. It allows to identify crystalline magnetic order as well as the magnitude and orientation of magnetic moments with high elemental specifity. Magnetic long range ordering leads to Bragg diffraction from which valuable structural magnetic information can be extracted. While the width of the Bragg peaks is coupled to the crystal quality or strain of magnetic lattices, their peak position and intensity allows to extract lattice parameters, content of the unit cell or the arrangement of magnetic moments. Monitoring these diffraction peaks under the influence of external fields, pressure, temperature variation, or growth thus enables a detailed $in$-$situ$ characterization of nanomagnetic systems. 
\\
During the last decades resonant x-ray diffraction and reflectometry were extensively applied to reveal the magnetic structure of various crystalline alloys \cite{Lang14, Hama12, Scag11} and superlattices \cite{Schla13, Toell95, Nagy02, Velt02, Hecker, Park}. Despite the growing need for characterization of extended patterned magnetic structures like spin ice \cite{Yu, Ross} or skyrmion thin film systems \cite{Castel, Wang, Branford} only a few resonant magnetic x-ray diffraction studies on such systems were carried out \cite{Kinane2006, Morgan2012}. Most of these studies were performed at the L$_{3}$ edges of Fe or Co and deal with the field dependent shape of nanostripe domain patterns \cite{Hellwig, Durr}. Using resonant soft x-ray diffraction, Chesnel $et$ $al.$ studied the field dependent magnetic configuration in a dipolar coupled multilayer grating \cite{Chesnel}. \\ 
Here we show that nuclear resonant surface diffraction provides unique insight into the magnetism of patterned nanostructures. Applicable to various alloys containing M\"ossbauer isotopes, the technique offers a significant number of features not accessible by other techniques. Nuclear diffraction allows to discriminate the resonantly scattered (delayed) x-rays from the dominating electronic (prompt) ones via time gating, enabling to detect a pure magnetic signal with very high signal-to-noise ratio. This yields superior sensitivity to detect magnetic superstructures in low-dimensional systems, weak contributions of magnetic order in a lattice or smallest variations of magnetic configurations under the influence of external stimuli. Synchrotron radiation in the hard x-ray regime offers access to magnetic correlation lengths down to the atomic scale. The large penetration depth enables depth dependent studies of lateral spin configurations in nanostructures with a height of a few ten nanometers. The vertical sensitivity can be uniquely brought to the atomic level by selectively doping the system with an isotopic probe layers at the depth of interest \cite{Ralf, Seb}.\\ 
While the distribution of nuclear resonant scattering in reciprocal space allows to identify lateral magnetic correlations in a lattice, its delayed temporal evolution after pulsed excitation contains site-selective information about magnetic moment orientations \cite{Ralf}, i.e., collecting the time dependent nuclear signal at different Bragg peak positions allows to selectively probe magnetic substructures in the sample. \\
Thus, nuclear time spectra are not only highly sensitive to the nanoscopic arrangement of magnetic moments but likewise sensitive to spin dynamics up to the GHz regime. Oscillation frequencies and precessional magnetic moment trajectories can be quantitatively analized \cite{Olaf, Lars15}. These features make nuclear resonant surface diffraction highly attractive for prospective studies of externally induced (thermal, current, magnetic field) lateral spin dynamics in magnetically patterned nano systems.\\
We use this novel scattering technique to investigate the magnetic structure of an array of nanostripes that can be tuned via application of an external magnetic field to assume a ferromagnetically or antiferromagnetically aligned state.
The nanostripe array, shown in Fig.\,\ref{setup}, prepared by electron-beam lithography, physical vapor deposition and lift-off processing. Nanowires consisting of Permalloy (Ni$_{80}$Fe$_{20}$, enriched to 95$\%$ in the M\"ossbauer isotope $^{57}$Fe) with a length of 20 $\mu$m, a height of 30 nm, a width of 175 nm and a lateral periodicity of 260 nm are arranged on a silicon wafer. The macroscopic sample consists of 500 nanostripe arrays of 20 $\mu$m x 400 $\mu$m each, forming a patch with a size of 10 mm x 0.4 mm. 
Domain-wall nucleation pads \cite{Yo00,Mi09} have been attached to every second nanostripe, thus reducing the coercive field of this set of stripes. This results in a two-step magnetic reversal of the lattice along the wire axis shown in Fig.\,\ref{setup}a. The magnetic hysteresis curve indicates a magnetic reversal with an intermediate zero net magnetization at around 40 mT which points to an antiferromagnetic configuration of the stripes at this field value. This is confirmed locally via magnetic force microscopy as shown in Fig.\,\ref{setup}b. In order to characterize the global magnetic state of this stripe array, we apply nuclear resonant diffraction in grazing incidence geometry.\\
The experiment was performed at the High-Resolution Dynamics beamline P01 at the synchrotron radiation source PETRA III at DESY in Hamburg \cite{P01}. The pulsed synchrotron radiation beam, resonantly  tuned to the 14.4 keV nuclear transition of $^{57}$Fe, is slightly focussed to match the pixel size of the x-ray area detector of 100 $\mu$m x 100 $\mu$m while still illuminating the full sample length. This allows to resolve lateral correlations in the sample ranging from the sub-nm to the $\mu$m regime. Non-resonant diffraction patterns are recorded in the beginning of the experiment to align the nanostripe lattice parallel to the beam (see supplemental information for details). The large extent of the diffraction pattern in reciprocal space, shown in Fig.\,\ref{setup}c, proves the high long-range structural quality of the nanostripe array.\\
To identify the magnetic structure of the lattice, non-resonant and nuclear resonant line scans are performed with an avalanche photo diode. The electronic signal is shown in Fig.\,\ref{fig:epsart}a which corresponds to a horizontal line cut through the 2D diffraction pattern at $\alpha_{out}=\alpha_{in}=0.2^\circ$. The angular positions of the diffraction peaks originate from the structural periodicity in the nanowire lattice with a lateral correlation length $\Lambda$ = 260 nm. In addition to the dominating nanowire diffraction peaks, weak peaks can be identified (black arrows) which are attributed to the pad structure with a doubled width and a doubled lattice parameter compared to the nanostripes. Fig.\,\ref{fig:epsart}b shows the nuclear resonant signal of the sample in magnetic saturation ($B_{\text{ext}}$=-100 mT). The angular distribution of the nuclear diffraction pattern reproduces the electronic ones because the magnetic correlation length $\Lambda_{\text{mag}}$ is identical to the structural correlation length. The situation drastically changes when the magnetic reversal in the lattice takes place, see Fig.\,\ref{fig:epsart}c and Fig.\,\ref{fig:epsart}d. Strong additional diffraction peaks appear with the first order peak at half the angular position of the structural signal. The doubled lateral magnetic correlation length thus confirms an antiferromagnetic configuration in the nanostripe lattice. The variation of the magnetic superstructure peak intensity as function of the external field allows to identify the magnetic reversal and the saturation behavior in the sample. The contribution of antiferromagnetic order in the nanostripe lattice is monitored via the intensity of the 1/2 order Bragg peak at $\Psi$=0.01$^\circ$ during a field sweep along the wire axis. This magnetization curve is shown in Fig.\,\ref{fig:epsart}e and characterized by zero intensity coming from saturation and thus absence of antiferromagnetic order up to 20 mT. The subsequent evolution of the nuclear intensity is dominated by a broad peak around $B_{\text{ext}}$ = +40 mT during the reversal process which is in excellent agreement with plateau-like features in the hysteresis curve, confirming the beginning and completion of the field dependent reversal. It reveals that every change in the net magnetization as seen by the hysteresis curve is translated to an increase or reduction of antiferromagnetic order in the lattice. We can thus conclude a uniaxial reversal via a purely antiferromagnetic state over the whole sample and exclude coupled domain wall motions of adjacent nanostripes. \\
These strong superstructure peaks prove that resonant x-ray diffraction is highly appropriate to study lateral spin configurations in ordered nanomagnetic systems. The technique can be applied to investigate large area sample systems with fast access to lateral nano- and micromagnetic correlations, integrated over the area of the x-ray foot print. This constitutes a very powerful method for in-situ studies of nanomagnetic systems during growth, for sample systems under the influence of external fields, electrical currents or large temperature variations. \\
Since the diffraction process at such a planar grating is essentially equivalent to the diffraction from a layered superlattice (see Fig.\,\ref{fig:wide}), the data evaluation can be performed via the evaluation software CONUSS that has been developed for evaluation of nuclear resonant scattering data \cite{Stur00}. To simulate the horizontal diffraction pattern we replace the lateral Py nanostripe lattice by a Py multilayer structure with vacuum spacer layers. The ratio of layer thickness to layer period is chosen to reproduce the ratio of nanostripe width to lateral period in the lattice. The vertical dimensions are reduced by a factor of 100 to account for the limited penetration depth of the x-rays and to result in strong Bragg reflections.
Thus, a stack of 30 Py layers with vacuum spacer layers is used for simulation with a thickness $d$ of 1.7 nm and a vertical periodicity $\Lambda_{\text{elec}}$ of 2.6 nm. \\
A comparison of the measured diffraction data and the calculated reflectivities is shown in Fig.\,\ref{fig:wide}. The angular scale of the reflectivity is compressed by a factor of 50 to account for the reduced multilayer thicknesses and angular dependent $q$-values ($q_{\text{hor}}\sim \sin{\Psi}, q_{\text{vert}}\sim 2\,\sin{\alpha}$). The pad structure is not included in the reflectivity model. In spite of this simplified model the calculated electronic and the resonant reflectivity curves reproduce the overall shape of the measured diffraction patterns well and likewise the strong suppression of the 3/2 order magnetic Bragg peak. The suppression of the magnetic diffraction peak can thus be attributed to the ratio of lateral magnetic correlation length in the lattice to the width of the nanostripes which is close to 3:1. Note, that the angular width of the experimental diffraction peaks is predominantly caused by the divergence and size of the synchrotron beam.   \\
Up to now we analyzed the angular dependent nuclear diffraction pattern, which acts as a fingerprint of the lateral magnetic configuration in the permalloy nanostripe pattern. The pulsed excitation of the magnetically split nuclear levels and the coherent re-emission of the resonant photons leads to a characteristic temporal beat pattern of the nuclear decay that depends on the orientation of the magnetic moments relative to the wavevector of the incident photon. To take full advantage of this new scattering technique we thus quantitatively analyze also the temporal beat pattern. Being sensitive to magnetic and electric hyperfine fields at the nuclei, these 'time spectra' additionally offer access to the chemical state as well as to magnetization dynamics of selected Fourier components of a nanostructure lattice that have been identified by the nuclear Bragg peaks. Figure \ref{timespectra} shows two time spectra of the lattice in the antiferromagnetic configuration ($B_{\text{ext}}$ = +40 mT) collected at the angular position of the
specularly reflected beam (0$^{th}$ order) and at the position of the first magnetic Bragg peak (1/2-order). Both beat patterns drastically differ due to the relative phasing of the contributions from the two sets of stripes with
$\mathbf{k_0}\uparrow\uparrow\mathbf{m}$ and $\mathbf{k_0}\uparrow\downarrow\mathbf{m}$. This can already be described in a simple model, based on a kinematical approximation.
Assuming a spatial phase factor of $e^{i\mathbf{q}\cdot\mathbf{R}} =: e^{i\phi}$ between the contributions from the two sets of stripes, one arrives at the following expression for the temporal beat pattern (see supplementary material):
\begin{equation}
I(t) = |\tilde{F}_P(t)|^2\,\cos^2\left(\frac{\phi}{2}\right) +  |\tilde{F}_M(t)|^2\sin^2\left(\frac{\phi}{2}\right) 
\end{equation}
where $\tilde{F}_P = \tilde{F}_+ + \tilde{F}_-$ and $\tilde{F}_M(t) = \tilde{F}_+ - \tilde{F}_-$ with $\tilde{F}_+$ ($\tilde{F}_-$) describing the temporal beating of the two nuclear transitions with change of magnetic quantum number $\Delta$m = $+1$ ($\Delta$m $= -1$), belonging to the emission of right- (left-) circular polarization. For a description of the time spectra in 0$^{th}$ and 1/2-order, we assume spatial phases of $\phi = 0$ and $\phi = \pi$, respectively, to account for the corresponding diffraction phases from a grating. We obtain that $I_0(t) = |\tilde{F}_P(t)|^2$ and  $I_{1/2}(t) = |\tilde{F}_M(t)|^2$, which provides a qualitatively good agreement with the measured data over the first 70 ns of the measured time spectra (see supplement). This already qualitatively confirms an antiparallel magnetization alignment of the nanostripes.  For a quantitative description of the data, the full dynamical theory of nuclear resonant reflectivity from thin films \cite{RRtheo} has to be applied, as implemented in CONUSS. Following the analogy laid out in Fig.\,\ref{fig:wide} and assuming a multilayer in an antiferromagnetic configuration, the time spectrum in the 0$^{th}$ order of the nanostripe diffraction is obtained in the 1$^{st}$ order Bragg peak of the multilayer where the contributions from the two magnetic sublattices add in phase ($\phi = 0$). Likewise, the time spectrum in the 1/2 order Bragg peak of the nanostripe array in Fig.\,\ref{timespectra} results from the sublattices adding in antiphase ($\phi = \pi$), and is thus equivalent to the time spectrum calculated for the first superstructure peak of the antiferromagnetic multilayer.
In summary, we have shown that nuclear resonant surface diffraction is highly suitable to study lateral spin configurations in ordered magnetic nanostructures with an outstanding signal-to-noise ratio. To demonstrate this, we detect strong resonant, background-free superstructure diffraction peaks from an antiferromagnetically ordered lattice of nanostripes. A simple theoretical approach was applied to evaluate the angular dependent as well as the time differential nuclear diffraction signal. The technique allows for a fast access to magnetic periodicities in macroscopic, nanostructured sample systems as well as to their magnetization dynamics, rendering it highly attractive for in-situ studies in the presence of external stimuli acting on the magnetic spin system. The simultaneous analysis of the nuclear diffraction pattern together with the temporal beat pattern of the nuclear signal in the magnetic diffraction peaks offers unique access to lateral domain configurations in surface-ordered magnetic nanostructures (size, magnetization direction and lateral arrangement). The technique will significantly benefit from time-resolving x-ray area detectors which will drastically shorten the data aquisition times, thus enabling efficient studies even of isotopically non-enriched sample systems. \\
We acknowledge financial support from the
Deutsche Forschungsgemeinschaft via SFB 668 'Magnetism from the Single Atom to the Nanostructure' and via excellence cluster 'The Hamburg Centre for Ultrafast Imaging - Structure, Dynamics and Control
of Matter on the Atomic Scale'.
\end{document}